# GAUGE INTERACTIONS AND A QUANTUM AVATAR IN A MODEL WITH LIGHT CONE REFLECTION SYMMETRY

Alan Chodos*
*Haseltine Systems, 2181 Jamieson Ave., Suite 1606*
*Alexandria, VA 22314*

**Abstract:** We continue the development of a model of neutrinos that possesses a new symmetry, Light Cone Reflection, interchanging spacelike and timelike intervals. In this paper we introduce gauge interactions in which only the physical modes of the theory participate. We also find an avatar of the theory, involving spinors in two-dimensions that obey equations reproducing the spectrum of the original, four-dimensional theory, and we discuss the quantization of the avatar.

## Introduction

Two previous papers [1,2] introduced and developed a discrete symmetry, Light Cone Reflection (LCR), that interchanges timelike and spacelike intervals. LCR is defined using a preferred null vector, $n^\mu$, which means that we are working not with a fully Lorentz-invariant theory, but in the context of Very Special Relativity (VSR), formulated a decade ago by Cohen and Glashow [3].

The motivation for LCR is to provide a rationale for the suggestion, made three decades ago [4], that one or more species of neutrino might be tachyonic. One expects that an LCR-symmetric theory will naturally possess both tachyonic and non-tachyonic degrees of freedom.

We note in passing that the relationship of neutrino velocities to the speed of light remains an open question despite the OPERA debacle of a few years ago [5]. The claim of the OPERA experiment, had it been true, would have suggested neutrino mass parameters greatly in excess of their expected range; the fact that their result was, ultimately, withdrawn says nothing about the possibility that some neutrinos are tachyons with meV-scale mass parameters.

In reference [2], we formulated a specific example of an LCR-symmetric theory, whose degrees of freedom are fermionic fields. We found a "special coordinate system" in which the physical part of the solutions depends only on the two light-cone coordinates; one is free to add an essentially arbitrary, unphysical part that is proportional to $\gamma^\mu n_\mu$. We showed that the resulting spectrum is indeed symmetric between tachyonic and non-tachyonic modes.

In this paper, we shall extend the analysis in two directions. Using a gauge principle, we shall introduce interactions in which only the physical modes participate; and, capitalizing on the fact that the physical modes obey equations in only two variables, we shall construct a two-dimensional avatar for these modes whose quantum properties we then describe.

In one respect, the analysis will be less general than that of reference [2], in which we solved the wave equation in the presence of both Dirac and Majorana mass terms. Because the gauge principle we shall use depends on a symmetry that is violated by the Majorana mass terms, in this work we shall deal with the purely Dirac case.

In the remainder of this paper, we first provide a very brief introduction to VSR and LCR; the interested reader should consult the earlier papers for somewhat more detail, as well as for additional references on related work on tachyons. Then we describe a hitherto unremarked symmetry of the Lagrangian, which we gauge to generate the interaction. We show that, as expected, only the physical modes are coupled by this interaction.

Following another line of investigation, we find a set of first-order equations whose square yields the two-dimensional Klein-Gordon-like equations, derived in [2], that led to the spectrum for the physical modes. We exhibit a Lagrangian for these first-order equations, and construct the quantum theory, showing that there is a well-defined, Hermitian Hamiltonian, bounded below, that generates the (light-cone) time development of this theory.

**VSR and LCR**

In the maximal version of what they called Very Special Relativity, Cohen and Glashow suggested that perhaps Nature is not invariant under the full Lorentz group, but only under a subgroup that either leaves a specified null vector $n^\mu$ invariant, or else scales $n^\mu$ by a constant (the scaling corresponds to a boost in the $\vec{n}$ direction). They wrote a VSR-invariant wave equation that reproduces the dispersion formula $p^2 = m^2$, while respecting chiral symmetry. Depending on exactly how Lorentz invariance is broken to VSR, they discussed to what extent one might expect to detect the lack of Lorentz invariance in a theory of neutrinos.

Making use of the null vector $n^\mu$ that VSR provides, we define LCR as the coordinate transformation

$$x^\mu \to y^\mu = x^\mu - n^\mu x^2 / n \cdot x \ ,$$

which clearly satisfies $y^2 = -x^2$ .

In [2] we proposed a VSR- , LCR- , and chiral-invariant Lagrangian for two neutrino fields $\psi_a$ and $\psi_b$ (in [2] these fields were labeled $\psi_\pm$, but we shall use $\pm$ for another purpose below). Following Alvarez and Vidal [6], we supplement $\psi_a$ and $\psi_b$ with two auxiliary fields, $\rho$ and $\chi$, and write the Lagrangian

$$L = L_\psi + L_{\chi\rho} + L_M$$

where

$$L_\psi = i\bar\psi_a \gamma^\mu D_\mu \psi_a + i\bar\psi_b \gamma^\mu D_\mu \psi_b ,$$

$$L_{\chi\rho} = i[\bar\chi (n \cdot \partial)\rho + \bar\rho (n \cdot \partial)\chi] , \text{ and}$$

$$L_M = i\{M_1 \bar\chi \gamma^\mu n_\mu \psi_a + M_2 \bar\rho \psi_b - h.c.\} .$$

This is the same Lagrangian that was used in [2], with the change in notation mentioned above, and keeping only the Dirac mass terms. The LCR-covariant derivative $D_\mu$ is defined as

$$D_\mu = \partial_\mu - \partial_\mu \phi \, n \cdot \partial ,$$

where $\phi$ is an additional degree of freedom that satisfies the constraint $n \cdot \partial \phi = 1$. As a consequence, we find $D_\mu \phi = 0$ and $n^\mu D_\mu = 0$.

As discussed in [2], with suitable transformation properties for the various fields, this Lagrangian is VSR-, LCR-, and chiral-invariant.

**Gauge Interaction**

What was not discussed in [2] is that L possesses additional symmetries, generated by

$$t_1 = in_\mu \gamma^\mu \text{ and } t_2 = \gamma_5 n_\mu \gamma^\mu$$

acting on $\psi_a$ and $\psi_b$ independently. Together with the chiral transformation generated by $t_3 = \frac{1}{2}\gamma_5$, these obey the algebra of rotations and translations in a plane.

More explicitly, let $\mathcal{M}(\alpha, \beta)$ be the matrix

$$\mathcal{M}(\alpha, \beta) = \exp\{\alpha n_\mu \gamma^\mu + i\beta \gamma_5 n_\mu \gamma^\mu\} = 1 + \alpha n_\mu \gamma^\mu + i\beta \gamma_5 n_\mu \gamma^\mu$$

where we have used the fact that $(n_\mu \gamma^\mu)^2 = 0$. Then L is invariant under the set of transformations

$$\psi_a \to \mathcal{M}(\alpha_a, \beta_a)\psi_a \; ; \quad \psi_b \to \mathcal{M}(\alpha_b, \beta_b)\psi_b \; ;$$

$$\chi \to \mathcal{M}(\alpha_b, \beta_b)\chi \; ; \quad \text{and} \quad \rho \to \mathcal{M}(-\alpha_b, \beta_b)\rho \; ,$$

where the $\alpha's$ and $\beta's$ are real constants. The property $n^\mu D_\mu = 0$ is crucial in establishing this invariance.

Our goal now is to gauge these symmetries. We let the $\alpha's$ and $\beta's$ be functions of the coordinates, so that L is no longer invariant under the above transformations, but undergoes the changes

$$\Delta L_\psi = i\overline{\psi}_a \gamma^\mu D_\mu (\alpha_a + i\beta_a \gamma_5) n_\nu \gamma^\nu \psi_a + i\overline{\psi}_b \gamma^\mu D_\mu (\alpha_b + i\beta_b \gamma_5) n_\nu \gamma^\nu \psi_b$$

and

$$\Delta L_{\chi\rho} = i[\overline{\chi} \, n \cdot \partial(-\alpha_b + i\beta_b \gamma_5) \, n_\mu \gamma^\mu \rho + \overline{\rho} \, n \cdot \partial(\alpha_b + i\beta_b \gamma_5) n_\mu \gamma^\mu \chi] \; .$$

To compensate for these, we introduce a set of gauge fields $(A_\mu, B_\mu)$ to $L_\psi$ and $(\hat{A}, \hat{B})$ to $L_{\chi\rho}$ as follows:

$$L_\psi^{int} = \sum_{j=a,b} \left[ i\overline{\psi}_j \gamma^\mu n \cdot \gamma \psi_j g_j A_\mu + \overline{\psi}_j \gamma^\mu \gamma_5 n \cdot \gamma \psi_j \tilde{g}_j B_\mu \right] \phi$$

and

$$L_{\chi\rho}^{int} = i(\overline{\chi}(n \cdot \gamma)\rho - \overline{\rho}(n \cdot \gamma)\chi)g_b \hat{A} + (\overline{\chi}\gamma_5(n \cdot \gamma)\rho + \overline{\rho}\gamma_5(n \cdot \gamma)\chi)\tilde{g}_b \hat{B} \; .$$

To insure hermiticity of $L_\psi^{int}$ we require that $n^\mu A_\mu = n^\mu B_\mu = 0$.

Here the $g_j$ and $\tilde{g}_j$ are real coupling constants, and $\phi$ is the field that enters the definition of the covariant derivative $D_\mu$, and that satisfies $D_\mu \phi = 0$ as well as the constraint $n \cdot \partial \phi = 1$. The reason for including the factor of $\phi$ will emerge below.

To achieve invariance, we must choose

$\alpha_j = g_j \, \alpha(x)$ and $\beta_j = \tilde{g}_j \, \beta(x)$, and let

$$A_\mu \to A_\mu - \frac{1}{\phi} D_\mu \alpha \; ; \quad B_\mu \to B_\mu + \frac{1}{\phi} D_\mu \beta \; ; \quad \hat{A} \to \hat{A} + n \cdot \partial \alpha \; ; \quad \text{and} \quad \hat{B} \to \hat{B} + n \cdot \partial \beta \; .$$

It is easy to see that under a chiral transformation,

$$\psi_j \to e^{i\theta\gamma_5}\psi_j \; ; \; \chi \to e^{i\theta\gamma_5}\chi \; ; \text{and} \; \rho \to e^{-i\theta\gamma_5}\rho$$

$(A_\mu, B_\mu)$ and $(\hat{A}, \hat{B})$ transform as chiral doublets, provided $g_a \tilde{g}_b - g_b \tilde{g}_a = 0$.

It is possible, although not necessary, to use the gauge freedom to completely gauge away $\hat{A}$ and $\hat{B}$, leaving a residual gauge symmetry in which $\alpha$ and $\beta$ are constrained to be independent of the light-cone variable $v$, for which $n \cdot \partial = 2 \partial/\partial v$.

For definiteness, we follow [1] and [2] by choosing our coordinates so that

$n^\mu = (1,0,0,1)$, $n_\mu = (1,0,0,-1)$.

In [2] we introduced a convenient spinor basis $\{|b_j>\}$, $j = 1,...,4$, given explicitly by

$|b_1> = \begin{pmatrix} \uparrow \\ \uparrow \end{pmatrix}$; $|b_2> = \begin{pmatrix} \downarrow \\ \downarrow \end{pmatrix}$; $|b_3> = \begin{pmatrix} \uparrow \\ -\uparrow \end{pmatrix}$; $|b_4> = \begin{pmatrix} \downarrow \\ -\downarrow \end{pmatrix}$,

where $\uparrow$ and $\downarrow$ denote eigenstates of $\sigma_z$ with eigenvalues +1 and -1 respectively. $|b_1>$ and $|b_4>$ are annihilated by $n \cdot \gamma$. $|b_1>$ and $|b_2>$ are eigenstates of $\gamma_5$ with eigenvalue +1, whereas $|b_3>$ and $|b_4>$ are eigenstates of $\gamma_5$ with eigenvalue $-1$.

We can expand our fields $\psi_{a,b}$ in this basis:

$\psi_{a,b} = \sum_{j=1}^{4} f_j^{(a,b)} |b_j>$ .

We adopt the gauge condition $\hat{A} = \hat{B} = 0$, for which $L_{\chi\rho}^{int} = 0$, and insert the expansion into $L_\psi^{int}$. After some algebra, we obtain

$L_\psi^{int} = 4 \phi \sum_{k=a,b} \{Im(f_3^{(k)*} f_2^{(k)})(g_k A_x + \tilde{g}_k B_y) - Re(f_3^{(k)*} f_2^{(k)})(g_k A_y - \tilde{g}_k B_x)\}$ .

As must be the case, the interaction involves only $f_2$ and $f_3$, which, as was found in [2], are the variables that satisfy dynamical equations. The coefficients $f_1$ and $f_4$, which are not determined dynamically, are deemed to be unphysical, and hence should not participate in the interaction.

We close this section with a discussion of kinetic terms for the fields $A_\mu$ and $B_\mu$. As mentioned earlier, $n^\mu$ scales under a VSR boost in the $\vec{n}$ direction: $n^\mu \to q n^\mu$. In order both to preserve the constraint $n \cdot \partial \phi = 1$ and to leave the covariant derivative $D_\mu$ unchanged, we demand that $\phi \to q^{-1} \phi$. It follows from the definition of $L_\psi^{int}$ that $A_\mu$ and $B_\mu$ remain invariant. So a VSR and LCR invariant kinetic term for $A_\mu$ and $B_\mu$ is

$L_{AB} = K[F_{\mu\nu}^A F^{A\mu\nu} + F_{\mu\nu}^B F^{B\mu\nu}]$, where $F_{\mu\nu}^A = D_\mu A_\nu - D_\nu A_\mu$ and likewise for $B_\mu$.

K is a constant to be chosen. Since $[D_\mu, D_\nu] = 0$, $L_{AB}$ has the necessary invariance under the gauge transformations $A_\mu \to A_\mu - \frac{1}{\phi} D_\mu \alpha$ and $B_\mu \to B_\mu + \frac{1}{\phi} D_\mu \beta$. Because both $n^\mu D_\mu = 0$ and $n^\mu A_\mu = n^\mu B_\mu = 0$, we see that, with the explicit choice for $n^\mu$ that we have made, $L_{AB}$ reduces to

$$L_{AB} = 2K\big[(D_x A_y - D_y A_x)^2 + (D_x B_y - D_y B_x)^2\big].$$

So $L_{AB}$ involves only the same variables, $A_x, A_y, B_x$, and $B_y$, that appear in $L_\psi^{int}$.

**Quantum Avatar**

We introduce the light-cone variables

$u = n_\mu x^\mu = t - z$ and $v = t + z$,

taking $u$ to be the evolution parameter. In [2] we chose a "special coordinate system" defined by adding to $v$ a particular function of the other variables. In that system, which we denote with primes, the $f's$ are required to be independent of the transverse coordinates $x'$ and $y'$, and they satisfy the following differential equations in $u'$ and $v'$:

$$\lambda f_2^{(a)} - B f_2^{(b)} = 0$$

$$\lambda f_3^{(a)} - B f_3^{(b)} = 0$$

$$\lambda f_2^{(b)} - B^* f_2^{(a)} = 0$$

$$\lambda f_3^{(b)} - B^* f_3^{(a)} = 0 .$$

Here $\lambda = 2\partial_{u'}\partial_{v'}$, and $B = -M_1^* M_2$. These are the same equations as were found in [2], except that, as noted above, the Majorana mass terms have been set to zero, which has the effect of decoupling the equations for $f_2$ and $f_3$.

These are second-order equations, analogous to the Klein-Gordon equation that is obtained after squaring the Dirac equation, and indeed these equations determine the spectrum of $\lambda$ to be $\mp|B|$, with the upper sign corresponding to ordinary particles and the lower sign to tachyons.

We shall exhibit a set of first-order, Dirac-like, equations in $u'$ and $v'$, whose square yields the equations above, and then we shall discuss their quantum properties. We refer to this two-dimensional system as the *avatar* of the full dynamical system, which is governed by the Lagrangian $L = L_\psi + L_{\chi\rho} + L_M$ defined in four-dimensional spacetime.

Consider the linear combinations

$$g_i^{(\pm)} = \mp \left(B^*/|B|\right) f_i^{(a)} + f_i^{(b)}, \quad i = 2,3.$$

These satisfy

$$\lambda g_i^{(\pm)} = \mp |B| g_i^{(\pm)} \equiv \mp 2\kappa^2 g_i^{(\pm)}.$$

Now define a pair of two-dimensional spinors

$$G^{(\pm)} = \begin{pmatrix} g_2^{(\pm)} \\ g_3^{(\pm)} \end{pmatrix}$$

and the matrix differential operator

$$\delta = i(\sigma_+ \partial_{u'} + \sigma_- \partial_{v'})$$

where $\sigma_+ = \begin{pmatrix} 0 & 1 \\ 0 & 0 \end{pmatrix}$ and $\sigma_- = \begin{pmatrix} 0 & 0 \\ 1 & 0 \end{pmatrix} = \sigma_+^\dagger$.

We see that $2\delta^2 = -\lambda$.

Our dynamical equations for the avatar therefore will be

$$\delta G^{(+)} = \kappa G^{(+)} \quad \text{and}$$

$$\delta G^{(-)} = i\kappa G^{(-)}.$$

$G^{(+)}$ describes slower than light particles, whereas $G^{(-)}$ describes tachyons.

To quantize this system, we need a suitable Lagrangian, from which we can then derive the canonical variables and impose anti-commutation relations.

Clearly the Lagrangian will have the form $L_0 = L^{(+)} + L^{(-)}$. For $L^{(+)}$ we take

$$L^{(+)} = \overline{G}^{(+)} \delta G^{(+)} - \kappa \overline{G}^{(+)} G^{(+)}$$

and for $L^{(-)}$

$$L^{(-)} = \overline{G}^{(-)} \delta G^{(-)} - i\kappa \overline{G}^{(-)} G^{(-)}$$

where $\overline{G}^{(+)} = G^{(+)\dagger}\sigma_x$ and $\overline{G}^{(-)} = G^{(-)\dagger}i\sigma_y$.

$L_0$ has the necessary hermiticity property, and yields the desired equations of motion.

It can be rewritten

$$L_0 = \tfrac{i}{2} G^{(+)\dagger}[(1 - \sigma_z)\partial_{u'} + (1 + \sigma_z)\partial_{v'}]G^{(+)}$$

$$+ \tfrac{i}{2} G^{(-)\dagger}[(1 - \sigma_z)\partial_{u'} - (1 + \sigma_z)\partial_{v'}]G^{(-)} - \kappa\left[G^{(+)\dagger}\sigma_x G^{(+)} + G^{(-)\dagger}\sigma_y G^{(-)}\right].$$

We observe that $L_0$ is symmetric under the transformations

$$G^{(+)}(u',v') \to \exp\left\{i\frac{\pi}{4}\sigma_z\right\} G^{(-)}(u',-v')$$

and

$$G^{(-)}(u',v') \to \exp\left\{-i\frac{\pi}{4}\sigma_z\right\} G^{(+)}(u',-v') .$$

This symmetry, interchanging the role of the tachyonic and non-tachyonic degrees of freedom, is the avatar of Light Cone Reflection.

Taking $u'$ as the evolution parameter, we deduce from $L_0$ the canonical momenta

$$\pi^{(\pm)} = \tfrac{i}{2} G^{(\pm)\dagger}(1 - \sigma_z) = \begin{pmatrix} 0 \\ ig_3^{(\pm)\dagger} \end{pmatrix}.$$

This tells us that $g_2^{(\pm)}$ is a dependent variable, and that we must impose the canonical anti-commutation relations

$$\left\{g_3^{(\pm)\dagger}(u',v'_1), g_3^{(\pm)}(u',v'_2)\right\} = \delta(v'_1 - v'_2) .$$

The Hamiltonian density following from $L_0$ is

$$\mathcal{H}_0 = \sum_{\pm} G^{(\pm)\dagger}(1 - \sigma_z)\partial_{u'} G^{(\pm)} ;$$

using the equations of motion

$$i\partial_{u'}g_3^{(+)} = \kappa g_2^{(+)} ; \; \partial_{u'}g_3^{(-)} = \kappa g_2^{(-)} ; \; i\partial_{v'}g_2^{(+)} = \kappa g_3^{(+)} ; \; \partial_{v'}g_2^{(-)} = \kappa g_3^{(-)}$$

and integrating by parts, we can cast the Hamiltonian in the manifestly Hermitian form:

$$H' = \frac{\kappa}{2} \int dv' \left[ g_2^{(+)\dagger} g_3^{(+)} + g_3^{(+)\dagger} g_2^{(+)} - i g_2^{(-)\dagger} g_3^{(-)} + i g_3^{(-)\dagger} g_2^{(-)} \right] \equiv H^{(+)} + H^{(-)} .$$

It is straightforward to show, with the help of the canonical anticommutators for $g_3^{(\pm)}$, and the anticommutators involving $g_2^{(\pm)}$ that follow from the equations of motion, that $H'$ correctly encodes the evolution of the system in the variable $u'$.

In order to discuss the Fock space on which $H'$ acts, it is convenient, as usual, to transform to momentum space. We introduce the Fourier transform of $g_3^{(\pm)}$:

$$g_3^{(\pm)}(u', v') = \frac{1}{\sqrt{2\pi}} \int dk \, \tilde{g}^{(\pm)}(k) \exp{-i \, (k_0^{(\pm)} u' - k v')}$$

where $k_0^{(\pm)}(k) = \mp \left( \kappa^2 / k \right)$ .

We deduce that

$$\{\tilde{g}^{(\pm)\dagger}(k), \tilde{g}^{(\pm)}(k')\} = \delta(k - k') .$$

We also find that $g_2^{(\pm)}$ have the Fourier representations:

$$g_2^{(+)}(u'. v') = -\frac{\kappa}{\sqrt{2\pi}} \int dk \, \frac{1}{k} \, \tilde{g}^{(+)}(k) \exp{-i \, (k_0^{(+)} u' - k v')} \quad \text{and}$$

$$g_2^{(-)}(u', v') = -\frac{i\kappa}{\sqrt{2\pi}} \int dk \, \frac{1}{k} \, \tilde{g}^{(-)}(k) \exp{-i \, (k_0^{(-)} u' - k v')} .$$

We take the factor $\left( 1/k \right)$ in these integrands to denote the principal value.

The Hamiltonians $H^{(\pm)}$ can now be expressed as:

$$H^{(\pm)} = \mp \kappa^2 \int dk \, \frac{1}{k} \, \tilde{g}^{(\pm)\dagger}(k) \tilde{g}^{(\pm)}(k) .$$

We observe that $H^{(+)}$ is positive in the integration range $-\infty < k < 0$, but negative for $0 < k < \infty$, whereas the opposite holds for $H^{(-)}$. Hence we re-order the operators, throwing away infinite constants as we do so, to obtain the positive expressions

$$H^{(+)} = \kappa^2 \left[ -\int_{-\infty}^{0} dk \, \frac{1}{k} \tilde{g}^{(+)\dagger}(k) \tilde{g}^{(+)}(k) + \int_{0}^{\infty} dk \, \frac{1}{k} \tilde{g}^{(+)}(k) \tilde{g}^{(+)\dagger}(k) \right]$$

and

$$H^{(-)} = \kappa^2 \left[ -\int_{-\infty}^{0} dk \frac{1}{k} \tilde{g}^{(-)}(k)\tilde{g}^{(-)\dagger}(k) + \int_{0}^{\infty} dk \frac{1}{k} \tilde{g}^{(-)\dagger}(k)\tilde{g}^{(-)}(k) \right]$$

From these expressions we learn that the vacuum state is defined by the conditions

$\tilde{g}^{(+)}(k)|0> = \tilde{g}^{(-)\dagger}(k)|0> = 0$, $k < 0$, and

$\tilde{g}^{(+)\dagger}(k)|0> = \tilde{g}^{(-)}(k)|0> = 0$, $k > 0$.

The Hilbert space is then built on this vacuum in the usual way, taking the creation operators to be the Hermitian conjugates of the operators that annihilate the vacuum.

We close this section by asking how this compares to what we would have obtained had we quantized the system by taking the evolution parameter to be the familiar time, that is, $t' = \frac{1}{2}(u' + v')$.[1] If the theory were Lorentz-invariant we would expect all physical properties to be the same, but in the case of VSR this expectation may well prove false.

To proceed, we construct the operator $P'$ that generates translations in $v'$:

$$P' = -\int_{-\infty}^{0} kdk \left( \tilde{g}^{(+)\dagger}(k)\tilde{g}^{(+)}(k) - \tilde{g}^{(-)}(k)\tilde{g}^{(-)\dagger}(k) \right)$$
$$+ \int_{0}^{\infty} kdk \left( \tilde{g}^{(+)}(k)\tilde{g}^{(+)\dagger}(k) - \tilde{g}^{(-)\dagger}(k)\tilde{g}^{(-)}(k) \right).$$

We have written $P'$ in normal-ordered form, so that $P'|0> = 0$.

Consider the states $|k>_{\pm}$ that are generated by the single application to the vacuum of a creation operator of $+$ or $-$ type respectively. We find that

$$H'|k>_{\pm} = \frac{\kappa^2}{|k|}|k>_{\pm}$$

and

$$P'|k>_{\pm} = \pm|k| \, |k>_{\pm}$$

so that $H'P'|k>_{\pm} = \pm\kappa^2 |k>_{\pm}$, confirming that the + states are non-tachyonic, while the – states are tachyonic.

---

[1] Note, though, that $t'$ corresponds to "ordinary" time in the special coordinate system, so it is not necessarily the time as measured in the laboratory.

The operators that generate translations in ordinary time and space are

$$H = \frac{1}{2}(H' + P') \text{ and } P = \frac{1}{2}(H' - P').$$

Acting on the $|k>_\pm$ states, these have the spectra

$$E = \frac{1}{2}\left(\frac{\kappa^2}{|k|} \pm |k|\right)$$

and

$$p = \frac{1}{2}\left(\frac{\kappa^2}{|k|} \mp |k|\right).$$

We observe that $E^2 - p^2 = \pm\kappa^2$.

For the + states, E > 0 for all $k$, so that no further normal ordering is necessary – when restricted to the + sector, the vacuum state remains the same. For the – states, however, E is negative for $|k| > \kappa$, indicating that the role of creation and annihilation operators must be redefined to obtain a Hamiltonian that is bounded below. In other words, for the – sector, the operator $H$ defines a different Fock space from the one defined by the operator $H'$. Had we been working with a Lorentz-invariant theory, this would be a signal that the existence of tachyons triggers a spontaneous breakdown of that symmetry.

**Conclusions**

The gauge interactions we have constructed have the desired property that only the physical modes participate. Assuming that our fields $\psi_{a.b}$ describe neutrinos, we would also like to introduce the interactions ascribed to them by the standard model. That remains for future consideration.

It would be interesting to examine the interactions at the level of the avatar, which might be a productive arena in which to study their properties. A more general question is to what extent the quantum properties of the avatar mirror those of the original four-dimensional model from which the avatar was derived.

Finally, an important issue raised in [2] has not been directly addressed in this paper, but its resolution is crucial to understanding how (or even whether) this model fits into the rest of physics. The "special coordinates" $x'^\mu$ that we have used were defined in [2] relative to the usual laboratory coordinates $x^\mu$ by the transformation

$$u' = u\,;\ x' = x\,;\ y' = y\,;\ v' = 2\phi\,,$$

where, as we have noted above, $\phi$ satisfies the constraint $2(\partial/\partial v)\phi = 1$, so that it can be written as $\frac{1}{2}v + h(u, x, y)$. Technically, $\phi$, or, more precisely, $h$, is a dynamical variable, not a coordinate, so the coordinate transformation above is a little unusual. We would like to know exactly how to determine $\phi$. Without that knowledge, one cannot reliably relate results obtained in the special coordinate system to the properties that can be measured in the laboratory.